# On the Residual-based Neural Network for Unmodeled Distortions in Coordinate Transformation

**Vinicius Francisco Rofatto[a,b(*)]**, **Luiz Felipe Rodrigues de Almeida[a]**,

**Marcelo Tomio Matsuoka[a,b]**, **Ivandro Klein[c,d]**, **Mauricio Roberto Veronez[e]**,

**Luiz Gonzaga Da Silveira Junior[e]**

[a]Institute of Geography, Geosciences and Collective Health – IGESC, Federal University of Uberlandia (UFU), Monte Carmelo Campus, 38500-000, Minas Gerais, Brazil.

[b]Graduate Program in Agriculture and Geospatial Information, Federal University of Uberlandia (UFU), Monte Carmelo Campus, 38500-000, Minas Gerais, Brazil.

[c]Department of Civil Construction, Federal Institute of Education, Science and Technology of Santa Catarina (IFSC), Florianópolis Campus, 88020-300, Florianópolis, Santa Catarina, Brazil.

[d]Graduate Program in Geodetic Sciences, Sector of Earth Sciences, Federal University of Paraná (UFPR), 81531-990, Curitiba, Paraná, Brazil.

[e]Advanced Visualization & Geoinformatics Lab-VizLab, Unisinos University, São Leopoldo 93022-750, Rio Grande do Sul, Brazil.

*Institute of Geography, Geosciences and Collective Health – IGESC, Federal University of Uberlandia (UFU), Monte Carmelo Campus, 38500-000, Minas Gerais, Brazil. E-mail address: vfrofatto@gmail.com (V.F. Rofatto). ORCID: https://orcid.org/0000-0003-1453-7530

## Abstract

Coordinate transformation models often fail to account for nonlinear and spatially dependent distortions, leading to significant residual errors in geospatial applications. Here we propose a residual-based neural correction (RBNC) strategy, in which a neural network learns to model only the systematic distortions left by an initial geometric transformation. By focusing solely on residual patterns, RBNC reduces model complexity and improves performance, particularly in scenarios with sparse or structured control point configurations. We evaluate the method using both simulated datasets—with varying distortion intensities and sampling strategies—and real-world image georeferencing tasks. Compared with direct neural network coordinate converters and classical transformation models, RBNC delivers more accurate and stable results under challenging conditions, while maintaining comparable performance in ideal cases. These findings demonstrate the effectiveness of residual modelling as a lightweight and robust alternative for improving coordinate transformation accuracy.

# 1. Introduction

Coordinate system transformation is a routine task in engineering and geoscience applications. Traditional models—such as isogonal, affine, and projective transformations—are widely used due to their analytical simplicity and computational efficiency [1,2]. However, these models are often misspecified because they fail to account for systematic distortion effects that are nonlinear, spatially dependent, and difficult to represent parametrically [3,4]. Such distortions typically appear in the residuals from least-squares adjustments and expose unmodelled structures that reduce transformation accuracy, particularly when datasets are heterogeneous or poorly conditioned [5].

In recent years, Artificial Neural Networks (ANNs) have emerged as flexible alternatives for coordinate transformation, with applications in geodetic reference frame conversion [6,7], image georeferencing [8], and three-dimensional point cloud registration [9]. Several studies have applied ANNs to perform direct transformations between coordinate systems, including conversions from geodetic to Cartesian coordinates [7], and global-to-local mappings that employ radial basis function networks with K-fold cross-validation [10]. More recently, Rofatto et al. (2023) [11] demonstrated that repeated leave-one-out cross-validation (RLOOCV) enables the estimation of pointwise uncertainty in neural network-based transformations.

Despite these advancements, most approaches employ neural networks as direct transformation models, where input coordinates in the source system are mapped directly to those in the target system. This structure requires the network to learn the entire transformation process simultaneously, including both the geometric relationship and the distortion effects. This dual requirement increases model complexity and tends to reduce generalization capability, especially under spatially varying distortions or when control points are sparse or unevenly distributed.

To address these limitations, this study proposes the Residual-Based Neural Correction (RBNC) approach. This method applies a geometric transformation first and then uses a neural network to model the residuals that remain. These residuals represent structured, spatially dependent components not explained by the analytical model. Rather than treating them as random noise, the method interprets them as informative signals that reflect model inadequacies and unmodelled effects. By using the residuals as learning targets, the neural network concentrates solely on capturing distortion patterns instead of performing the full transformation.

Although some recent studies have incorporated ANNs to compensate for distortions left by classical transformation models (e.g., [5,12]), they have not yet examined how transformation quality depends on control point configuration, sampling strategy, or network architecture. This study evaluates the proposed method under different conditions that include distortion complexity, sampling strategy (random vs. systematic), and network size (number of neurons), using both simulated and real-world datasets. The simulated experiments provide controlled scenarios with weak, moderate, and strong distortions. The real-world analysis focuses on image georeferencing tasks, where the transformation between image coordinates and object-space coordinates is evaluated under varying sample sizes and control point distributions.

Here, we also investigate the circumstances under which RBNC outperforms a direct Neural Network Coordinate Converter (NNCC), as well as situations in which both approaches provide equivalent results. The traditional geometric transformation model is also included as a baseline for comparison. This framework allows for an analysis of the practical utility of residual-based modelling and its relationship with data sampling and network structure in coordinate transformation tasks.

The remainder of this paper is structured as follows. Section 2 presents the Residual-Based Neural Correction (RBNC) method in detail. Section 3 describes the experimental setup, including both the simulated and real-world datasets. The simulated experiments explore distinct distortion patterns—barrel and pincushion—across weak, moderate, and strong levels. The real-world application focuses on image georeferencing using non-metric imagery, where image coordinates are transformed into object-space coordinates under different sample sizes and control point configurations, mimicking operational constraints in field applications. In this section, performance comparisons are drawn between RBNC, a direct Neural Network Coordinate Converter (NNCC), and a traditional geometric model (GeoTransf), offering insight into the conditions under which each method performs best. Section 4 concludes the study by summarizing key findings, discussing practical implications, and suggesting future directions.

## 2. Proposed Method: RBNC – Residual-Based Neural Network Correction

The proposed method is displayed in Fig. 1.

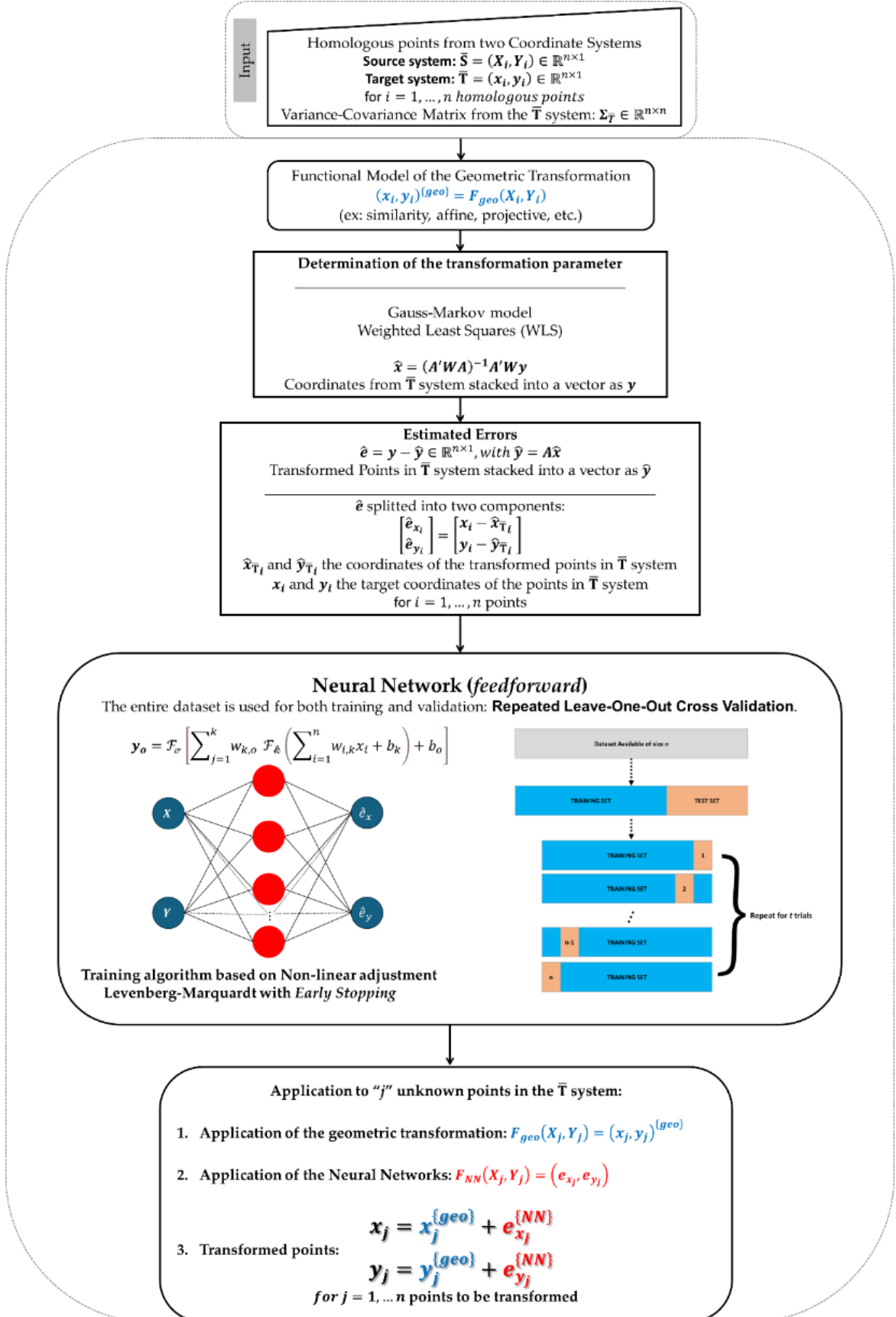

**Fig. 1.** Flowchart of RBNC – Residual-Based Neural Network Correction

The procedure can be summarized in the following stages (Figure 1):

1. Homologous points (control points) from two coordinate systems are required, namely the source system $\bar{S} := (X_i, Y_i)$ and the target system $\bar{T} := (x_i, y_i)$, for $i = 1, \ldots, n$. The minimum number of required control points depends on the transformation model employed. For instance, in the case of the 2D Similarity/Helmert model, which involves four transformation parameters—one rotation angle, one scale factor, and two translations)—at least three well-distributed points are necessary. This ensures that the least squares adjustment meets the minimal redundancy requirements, which is essential for uncertainty estimation. A variance-covariance matrix $\Sigma_{\bar{T}} \in \mathbb{R}^{n \times n}$ from the target system can also be included to weight the observations appropriately.

2. A functional model $(x_i, y_i)^{(geo)} = F_{\text{geo}}(X_i, Y_i)$ (e.g., similarity/Helmert, affine, projective, or polynomial transformation) is defined to describe the geometric relationship between source and target systems. Transformation parameters are estimated using a Gauss–Markov model with Weighted Least Squares (WLS), yielding transformed coordinates $(\hat{x}_{\bar{T}_i}, \hat{y}_{\bar{T}_i})$.

3. The least-squares residuals are then computed as the difference between the actual coordinates in the target system $(x_i, y_i)$ and the transformed coordinates obtained from the geometric model $(\hat{x}_{\bar{T}_i}, \hat{y}_{\bar{T}_i})$, i.e., $\hat{e}_{x_i} = x_i - \hat{x}_{\bar{T}_i}$ and $\hat{e}_{y_i} = y_i - \hat{y}_{\bar{T}_i}$ for $i = 1, \ldots, n$. Residuals represent the estimated coordinate errors in the target system with respect to the transformed values from the source system.

4. A feedforward neural network is then incorporated at this stage to predict the residuals in the target system based on the spatial position of each point in the source system. The network approximates a nonlinear function that maps $\bar{S} := (X_i, Y_i)$ to $(\hat{e}_{x_i}, \hat{e}_{y_i})$, i.e., $F_{NN}(X_i, Y_i) = (\hat{e}_{x_i}, \hat{e}_{y_i})$.

    i. Training follows the configuration proposed by Rodrigues et al. (2022), using the Levenberg–Marquardt optimization algorithm with early stopping to avoid overfitting. However, in this study, after several trials, the hyperbolic tangent was chosen to be as the activation function. Input and output data were standardized by scaling their minimum and maximum values to –1 and 1, respectively, as described in Equation 5 of [13].

ii. The network is trained and evaluated using a Repeated Leave-One-Out Cross-Validation (RLOOCV) strategy. In each iteration, one point is excluded from the dataset and used as an external validation sample (test point), while the remaining $n - 1$ points form the training set. From these, 90% are randomly selected for weight neural network adjustment, and the remaining 10% are used for internal validation to monitor the learning process. Early stopping is applied based on the internal validation error to prevent overfitting. Training was interrupted when the validation error failed to improve after six consecutive iterations. This procedure is repeated for a predefined number of trials to ensure stability and robustness of the results.

   a. In this study, 100 trials were conducted, following the recommendations of [11]. The neural network was trained using the command-line functionality of the Neural Network Toolbox in MATLAB R2024b.
   
   b. The RLOOCV procedure generates multiple prediction realizations for each test sample. This repetition enables the assessment of the variability associated with the training process. Both the mean and the median of the predicted values are computed as potential final estimates for each coordinate.

5. For any new point $(X_j, Y_j)$ in the source system $\overline{S}$, the transformation is performed in two steps. First, the geometric model is applied to obtain the transformed coordinates $(\hat{x}_j, \hat{y}_j)^{(geo)} = F_{\text{geo}}(X_j, Y_j)$. The ensemble of trained neural networks is used to estimate the residuals at that location by computing the mean and/or median of the predicted values: $F_{NN}(X_j, Y_j) = (\hat{e}_{x_j}, \hat{e}_{y_j})$. The implications of using the mean or the median as predictors will be discussed later.

6. The final coordinates in the target system are then computed by adding the predicted residuals to the geometric output:

$$(\hat{x}_j, \hat{y}_j) = (\hat{x}_j, \hat{y}_j)^{\{geo\}} + (\hat{e}_{x_j}, \hat{e}_{y_j})^{NN} \tag{1}$$

It is important to highlight that the neural network is only applied if, and only if, the least-squares residual pattern is not random. In cases where residuals exhibit no systematic structure, the use of a neural network is not justified. Therefore, the method assumes that model misspecification arises from the absence of parameters capable of describing systematic distortion effects, for which the selected geometric transformation fails to account.

## 3. Experimental Setup and Results

To assess the performance and applicability of the proposed method, two experiments were conducted: one with simulated data, designed to evaluate the method under controlled conditions with defined transformation parameters, distortions, and noise; and another with real-world measurements, providing a realistic assessment in scenarios with natural measurement noise and unmodeled distortions. Both experiments involved dataset construction, transformation models, and an evaluation strategy comparing three approaches: (i) GeoTransf, applying only the geometric transformation without correction; (ii) the proposed method (RBNC), which models residual distortions using a neural network; and (iii) NNCC, a direct coordinate conversion via neural network using source system coordinates as input and target system coordinates as output.

### 3.1 Simulated Experiment: Projective Transformation and Radial Distortions

This experiment was designed to evaluate the proposed method under conditions involving projective transformations and systematic distortions. The transformation between coordinate systems was first defined by a general projective model, constructed as a composition of linear and nonlinear effects. Additional radial distortions were then simulated to represent typical systematic errors in computer vision applications [14].

A regular grid of points was generated over a square region measuring 100 × 100 meters, with 10-meter spacing between adjacent points (Fig. 2), resulting in a total of $n = 121$ points.

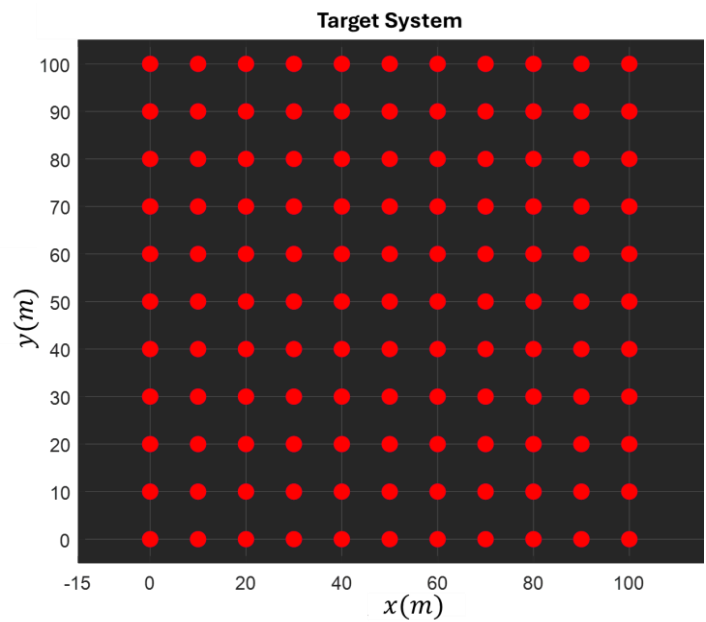

**Fig. 2.** Spatial distribution of the points in the target coordinate system.

These coordinates are referenced in the target system as $\bar{T} := (x_i, y_i)$, for $i = 1,\ldots,121$ points, which can be represented in homogeneous form as:

$$P_{\bar{T}_i} = \begin{pmatrix} x_i \\ y_i \\ 1 \end{pmatrix}, \quad i = 1,\ldots,121 \tag{2}$$

To simulate the source system $\bar{S} := (X_i, Y_i)$, each point was transformed using a composite projective matrix $H$, defined as the product of four transformation matrices, as follows:

1) Similarity transformation $H_1$: rotation, isotropic scale, and translation:

$$H_1 = \begin{pmatrix} s\cos\theta & -\sin\theta & t_x \\ s\sin\theta & \cos\theta & t_y \\ 0 & 0 & 1 \end{pmatrix} \tag{3}$$

with $s = 1.3$, $\theta = 25°$, $t_x = 3000m$ and $t_y = 5000m$.

2) Shear transformation $H_2$:

$$H_2 = \begin{pmatrix} 1 & k & 0 \\ 0 & 1 & 0 \\ 0 & 0 & 1 \end{pmatrix}, \text{ with } k = 0.08 \tag{4}$$

3) Anisotropic scaling $H_3$:

$$H_3 = \begin{pmatrix} \lambda & 0 & 0 \\ 0 & 1/\lambda & 0 \\ 0 & 0 & 1 \end{pmatrix}, \text{ with } \lambda = 1.15 \tag{5}$$

4) Elation (projective distortion) $H_4$:

$$H_4 = \begin{pmatrix} 1 & 0 & 0 \\ 0 & 1 & 0 \\ v_1 & v_2 & v \end{pmatrix}, \text{ with } v_1 = 0.0006, v_2 = -0.0004, v = 1 \tag{6}$$

The complete transformation matrix $H$ is then computed as:

$$H = H_1 \cdot H_2 \cdot H_3 \cdot H_4 = \begin{pmatrix} \lambda s\cos\theta + t_x v_1 & \frac{ks\cos\theta - \sin\theta}{\lambda} + t_x v_2 & t_x v \\ \lambda s\sin\theta + t_y v_1 & \frac{ks\sin\theta + \cos\theta}{\lambda} + t_y v_2 & t_y v \\ v_1 & v_2 & v \end{pmatrix} \tag{7}$$

The homogeneous coordinates in the source system $(X_{hi}, Y_{hi})$ are then obtained as follows:

$$P_{\bar{S}_i} = H \cdot P_{\bar{T}_i} = \begin{pmatrix} X_{hi} \\ Y_{hi} \\ w_i \end{pmatrix}, \quad i = 1,\ldots,121 \tag{8}$$

The Cartesian coordinates $(X_i, Y_i)$ are recovered by normalization:

$$X_i = \frac{X_{hi}}{w_i}, \quad Y_i = \frac{Y_{hi}}{w_i} \quad (9)$$

where $w_i$ is the result of the third row of $\boldsymbol{H}$ acting on the coordinated homogenous target system $\boldsymbol{P}_{\bar{T}_i}$ in Equation (3). Specifically, due to the elation component, $w_i$ is given by:

$$w_i = v_1 x_i + v_2 y_i + v, \quad i = 1, \ldots, 121 \quad (10)$$

This normalization step ensures that each transformed point is correctly mapped into two-dimensional space. These coordinates are shown in Fig. 3.

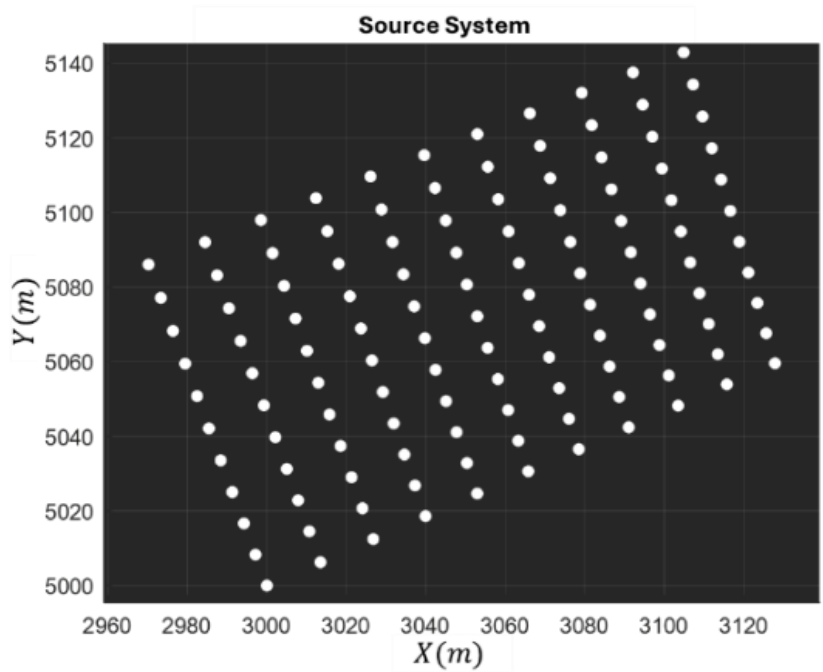

**Fig. 3.** Source system points generated by projective transformation only.

To simulate additional systematic distortions, a radial distortion model was applied to the simulated coordinates $(X_i, Y_i)$. The distorted coordinates $(X_i^d, Y_i^d)$ in source system were generated as:

$$\begin{aligned} X_i^d &= X_i + (X_i - X_c) \cdot K \cdot r_i^2 \\ Y_i^d &= Y_i + (Y_i - Y_c) \cdot K \cdot r_i^2 \end{aligned} \quad (11)$$

with

$$r_i^2 = (X_i - X_c)^2 + (Y_i - Y_c)^2, \quad i = 1, \ldots, 121 \quad (12)$$

The centre point $(X_c, Y_c)$ corresponds to the centroid of the coordinates in source system. The distortion coefficient $K$ controls both the magnitude and the type of distortion: when $K > 0$, the radial distortion is referred to as *pincushion distortion*; when $K < 0$, it is referred to as *barrel distortion*. Here, three levels of distortion were tested: Strong: $|K| = 5 \times 10^{-5}$ (Fig. 4a. – barrel distortion; Fig. 4d. – *pincushion distortion*); Moderate: $|K| = 3 \times 10^{-5}$ (Fig. 4b. – barrel distortion; Fig. 4e. – *pincushion distortion*); and Weak: $|K| = 1 \times 10^{-5}$ (Fig. 4c. – barrel distortion; Fig. 4f. – *pincushion distortion*).

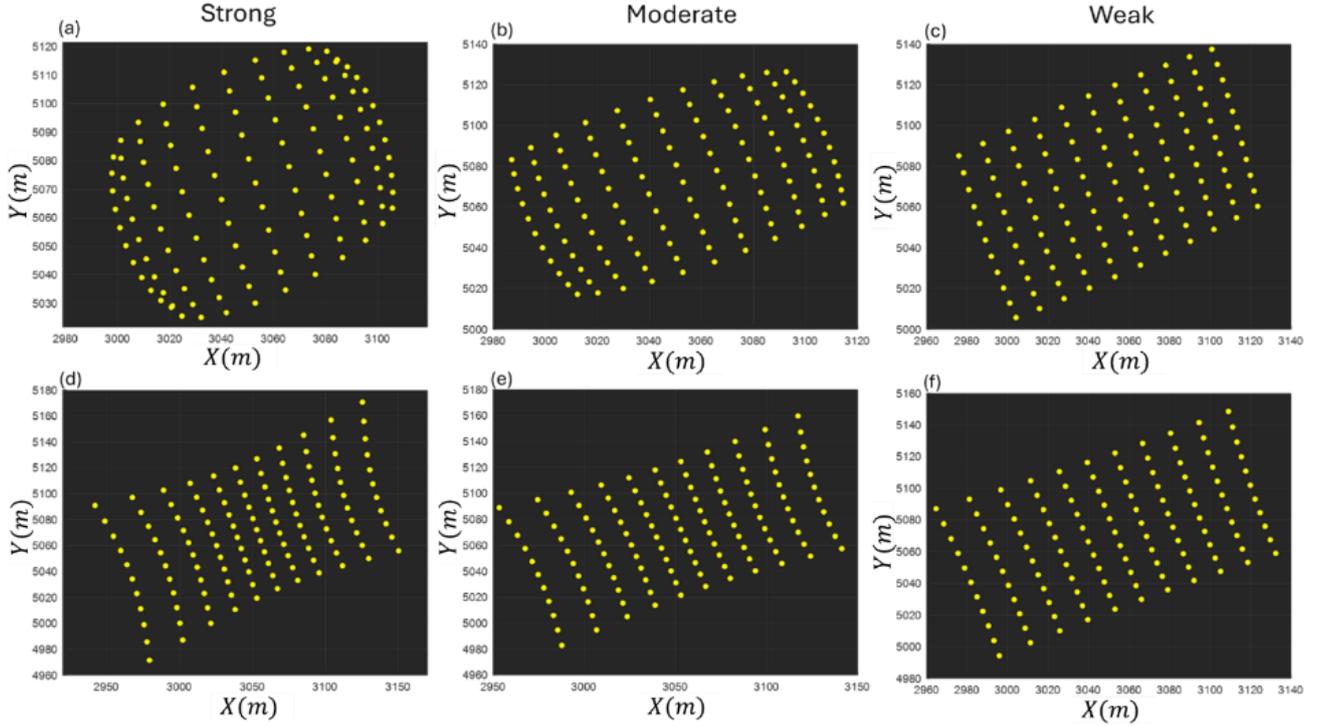

**Fig. 4**. Source system coordinates obtained by applying simulated radial distortions to the projectively transformed points from the target system. Top row (a–c): barrel distortion; bottom row (d–f): pincushion distortion. Each column represents a different distortion level: strong, moderate, and weak.

To simulate observational uncertainty and to characterize the coordinates as random variables, zero-mean Gaussian noise was independently added to both systems. The noise was sampled from a bivariate normal distribution $\mathcal{N}(\boldsymbol{\mu}, \boldsymbol{\Sigma})$, where $\boldsymbol{\mu}$ is the mean vector and $\Sigma$ is the covariance matrix. For the distorted source system $\overline{S}$, the standard deviation was 0.01 m, and for the target system $\overline{T}$, it was 0.001 m. The noisy coordinates are expressed as:

$$(X_i, Y_i) = (X_i^d, Y_i^d) + \mathcal{N}\left([0,0], \begin{bmatrix} 0.01^2 & 0 \\ 0 & 0.01^2 \end{bmatrix}\right), i = 1, \ldots, 121 \qquad (13)$$

$$(x_i, y_i) = (x_i, y_i) + \mathcal{N}\left([0,0], \begin{bmatrix} 0.001^2 & 0 \\ 0 & 0.001^2 \end{bmatrix}\right), i = 1, \ldots, 121 \qquad (14)$$

Following the generation of simulated data, the proposed and comparative methods (GeoTransf and NNCC) were applied to assess their performance under various distortion scenarios.

In the GeoTransf strategy, the geometric transformation model was based on the projective formulation defined by the Direct Linear Transformation, as described by [15, 16], as follows:

$$x_i = a_1 X_i + a_2 Y_i + a_3 - x_i(a_7 X_i + a_8 Y_i), i = 1, \ldots, 121 \tag{15}$$

$$y_i = a_4 X_i + a_5 Y_i + a_6 - y_i(a_7 X_i + a_8 Y_i), i = 1, \ldots, 121 \tag{16}$$

being $a_1, a_2, a_3, a_4, a_5, a_6, a_7, a_8$ the unknown transformation parameters.

The direct projective transformation can be formulated within the Gauss–Markov model framework, as follows [17]:

$$\boldsymbol{y} = A\boldsymbol{x} + \boldsymbol{\varepsilon}, \quad E[\varepsilon] = 0, \quad \text{Cov}[\varepsilon] = \Sigma_y \tag{17}$$

with:

- $\boldsymbol{y} \in \mathbb{R}^{2n \times 1}$: observation vector (coordinates in the target system), composed by:

$$\boldsymbol{y} = \begin{pmatrix} x_1 \\ y_1 \\ x_2 \\ y_2 \\ \vdots \\ x_n \\ y_n \end{pmatrix} \tag{18}$$

- $\boldsymbol{x} \in \mathbb{R}^{u \times 1}$: vector of $u = 8$ unknown parameters to be estimated:

$$\boldsymbol{x} = \begin{pmatrix} a_1 \\ a_2 \\ a_3 \\ a_4 \\ a_5 \\ a_6 \\ a_7 \\ a_8 \end{pmatrix} \tag{19}$$

The functional model that relates the unknown parameter vector $\boldsymbol{x}$ to the target coordinates in $\boldsymbol{y}$ through the design matrix $\boldsymbol{A} \in \mathbb{R}^{2n \times 8}$ was formulated by applying a scalar multiplier of $10^3$ in $c$ and $f$ (translation parameters) for numerical stability, as follows:

$$A = \begin{pmatrix} X_1 & Y_1 & 10^3 & 0 & 0 & 0 & -X_1x_1 & -Y_1x_1 \\ 0 & 0 & 0 & X_1 & Y_1 & 10^3 & -X_1y_1 & -Y_1y_1 \\ X_2 & Y_2 & 10^3 & 0 & 0 & 0 & -X_2x_2 & -Y_2x_2 \\ 0 & 0 & 0 & X_2 & Y_2 & 10^3 & -X_2y_2 & -Y_2y_2 \\ \vdots & \vdots & \vdots & \vdots & \vdots & \vdots & \vdots & \vdots \\ X_n & Y_n & 10^3 & 0 & 0 & 0 & -X_nx_n & -Y_nx_n \\ 0 & 0 & 0 & X_n & Y_n & 10^3 & -X_ny_n & -Y_ny_n \end{pmatrix} \quad (20)$$

- $\boldsymbol{\varepsilon} \in \mathbb{R}^{2n \times 1}$: vector of unknown residual to be estimated
- $\Sigma_y \in \mathbb{R}^{2n \times 2n}$: variance–covariance matrix of the observation vector
- $W = \Sigma_y^{-1} \in \mathbb{R}^{2n \times 2n}$: weight matrix.

With the matrices defined, the transformation parameters were estimated by applying the weighted least squares criterion. The solution is given by:

$$\hat{\boldsymbol{x}} = (A^\top \boldsymbol{W} A)^{-1} A^\top \boldsymbol{W} \boldsymbol{y} \quad (21)$$

After estimating the parameters, the adjusted observations were computed as:

$$\hat{\boldsymbol{y}} = A\hat{\boldsymbol{x}} \quad (22)$$

and the vector of the estimated errors (residual vector) was obtained from:

$$\hat{\boldsymbol{e}} = \boldsymbol{y} - \hat{\boldsymbol{y}} \quad (23)$$

These residuals represent the coordinate errors in the target system relative to the values estimated by the geometric transformation model.

To enable cross-validation, a Leave-One-Out Cross-Validation (LOOCV) strategy was adopted. In each iteration, one control point was removed from the dataset and treated as the test sample, while the remaining $n - 1$ points were used to estimate the transformation model. This process was repeated for all control points (homologous points). This approach ensured that the same test samples were used to evaluate the RBNC and NNCC methods, which allowed a fair and consistent comparison among all approaches.

It is important to highlight that, for computing prediction errors ($\bar{e}$)—i.e., errors associated with new observations not used in the adjustment (test samples)—it is not appropriate to apply the direct model given in equations (15) and (16), since these models depend on the coordinates from the target system, as also reflected in the design matrix $A$ (20). For this reason, the prediction step for the test set was performed using the classical projective transformation model, after estimating the parameters via Equation (21).

The predicted coordinates for the test point were obtained using the classical projective formulas:

$$\bar{x}_i^{\{geo\}} = \frac{\hat{a}X_i+\hat{b}Y_i+\hat{c}}{\hat{g}X_i+\hat{h}Y_i+1}, \quad \bar{y}_i^{\{geo\}} = \frac{\hat{d}X_i+\hat{e}Y_i+\hat{f}}{\hat{g}X_i+\hat{h}Y_i+1}, \text{ for } i = 1, \ldots, 121 \qquad (24)$$

The prediction errors were then computed by comparing the predicted values with the known coordinates from the target system:

$$\bar{e}_{x_i}^{\{geo\}} = \bar{x}_i^{\{geo\}} - x_i, \quad \bar{e}_{y_i}^{\{geo\}} = \bar{y}_i^{\{geo\}} - y_i, \text{ for } i = 1, \ldots, 121 \qquad (25)$$

In the Neural Network Coordinate Converter (NNCC) strategy, the network was configured to generate the nonlinear mapping from the source coordinate system to the target system. The input consisted of coordinates in the source system, while the output corresponded to the respective coordinates in the target system. The Levenberg Marquardt optimizer and the early stopping criteria followed the same configuration proposed by [11], that is, the same used in the RBNC method. The Repeated Leave-One-Out Cross-Validation strategy applied in the RBNC method was also adopted here, with 100 repetitions. Both the mean and the median of the predicted values were considered as potential final estimates for each coordinate. These outputs were subsequently used for performance comparison.

Thus, the prediction errors for NNCC and RBNC were computed in a similar way to Equation (25), with the following expressions:

$$\bar{e}_{x_i}^{(m)} = \bar{x}_i^{(m)} - x_i, \quad \bar{e}_{y_i}^{(m)} = \bar{y}_i^{(m)} - y_i, \quad \text{for } i = 1, \ldots, 121 \quad \text{with } m \in \{\text{NNCC,RBNC}\} \qquad (26)$$

Here, $\bar{x}$ and $\bar{y}$ represent either the mean or the median of the 100 prediction realizations obtained through the RLOOCV procedure. The resulting error can be computed as follows:

$$\bar{e}_{r_i}^{(m)} = \sqrt{\left[\bar{e}_{x_i}^{(m)}\right]^2 + \left[\bar{e}_{y_i}^{(m)}\right]^2} \text{ for } i = 1, \ldots, 121 \quad \text{with } m \in \{\text{NNCC,RBNC}\} \qquad (27)$$

Here, the resulting mean error was considered in our analysis and computed as:

$$\sum_{i=1}^{n} \frac{\bar{e}_{r_i}^{(m)}}{n} \text{ for } i = 1, \ldots, 121 \quad \text{with } m \in \{\text{NNCC,RBNC}\} \qquad (28)$$

A recurrent issue in neural network applications involves determining the appropriate number of neurons. This question was also explored in this study and will be further presented in the next subsection Results for Simulated Experiments.

## 3.2 Results for Simulated Experiments

The performance results for the simulated scenarios are summarized in Fig. 5. The experiments encompassed different distortion patterns, namely barrel and pincushion distortions, across varying intensities (strong, moderate, and weak). For both NNCC and RBNC, the final outcome was derived using two prediction schemes: the mean (denoted as NNCC-mean and RBNC-mean) and the median (NNCC-median and RBNC-median).

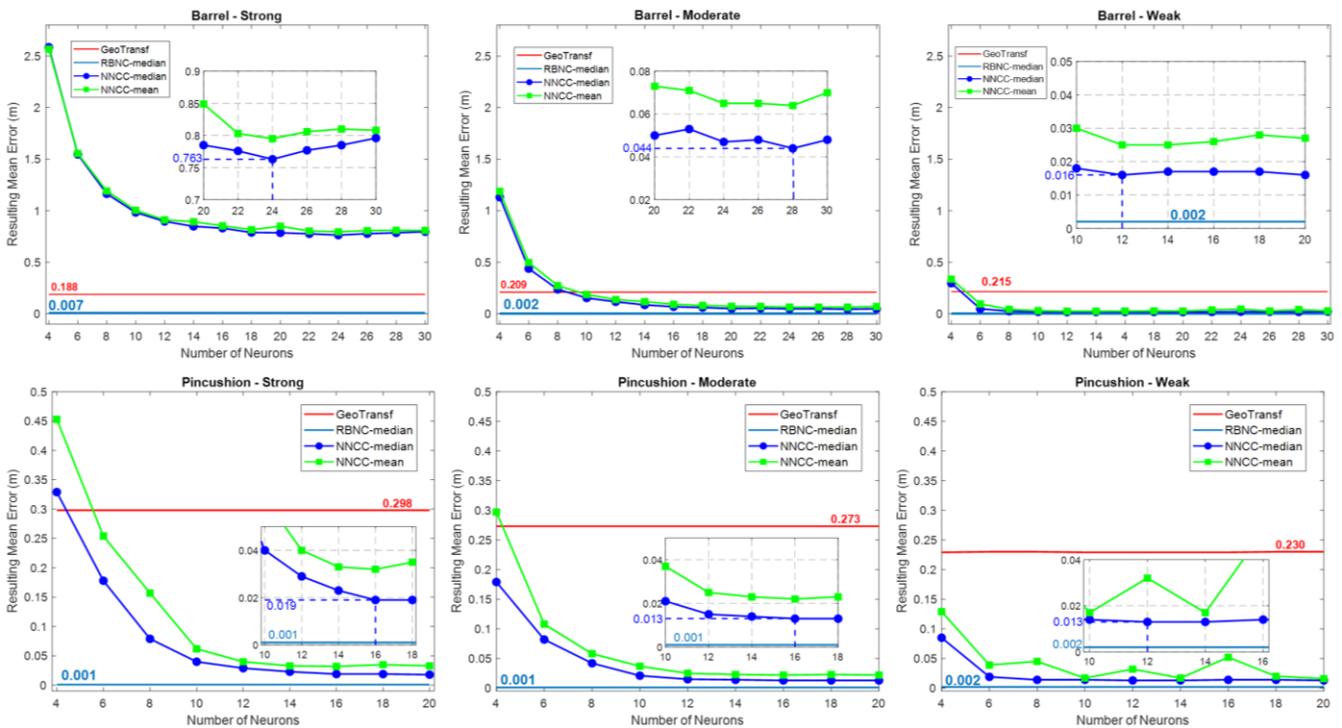

**Fig. 5.** Resulting mean error for NNCC and RBNC under different distortion patterns and numbers of neurons, alongside the GeoTransf method to support the analysis.

For the GeoTransf approach, the error behaviour reflects the misspecification of the transformation model. Although the method consistently yielded the resulting mean errors in the range of approximately 20 to 30 centimetres across scenarios, this outcome is expected since the model lacks sufficient parameters to capture the distortions. Incorporating additional terms could potentially enhance its ability to model complex geometric deformations. However, such improvements would require prior knowledge of the distortion pattern or the selection of a specific model to represent the residual structure, which is not a straightforward task.

The NNCC method, in contrast, exhibited sensitivity to both the distortion pattern and the number of neurons in the hidden layer. Although increasing the neuron count generally improved performance, excessive neurons led to diminishing returns, particularly in scenarios where the distortion caused clusters of points to overlap spatially at the edges of the grid (see the red ellipses in Fig. 6, strong barrel distortion). This overlap results in multiple input coordinates becoming nearly identical, which reduces the input diversity and compromises the network's capacity to generalize. This limitation is further evidenced by the spatial distribution of errors illustrated in Fig. 6 (right), where higher errors are concentrated precisely in these overlapping regions.

Moreover, the optimal number of neurons was found to depend on the type and intensity of the distortion (Fig. 5.). For scenarios of lower distortion intensity, as observed, 12 neurons were sufficient to reduce the resulting error to approximately 2 cm. Conversely, more severe distortions required a larger number of neurons: around 16 neurons for the pincushion scenario and approximately 28 neurons for the barrel distortion.

As previously mentioned, when severe overlap occurs, resulting in identical pairs of input attributes, the use of a neural network as a direct converter becomes unsuitable. Furthermore, between the two prediction strategies employed in the NNCC, the median consistently demonstrated greater robustness compared to the mean, particularly in scenarios with high distortion intensity. Nevertheless, even with this advantage, the RBNC method combined with the median predictor maintained superior overall performance.

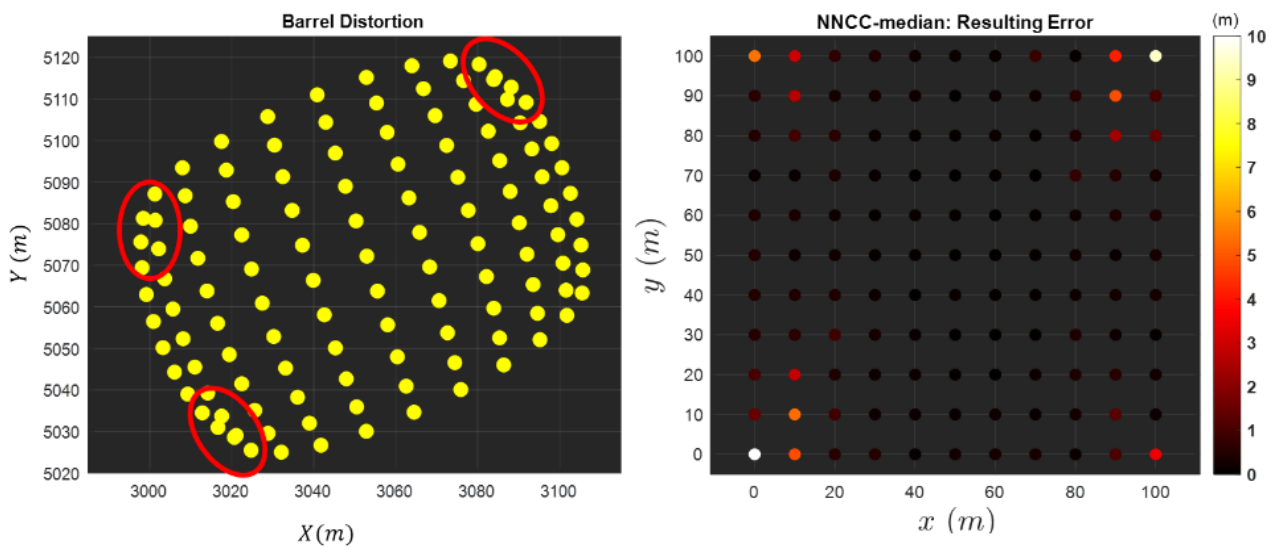

**Fig. 6.** Spatial overlap of input points, with highlighted regions in red (left), and resulting error distribution for the NNCC-median (right) under the strong barrel distortion scenario.

For RBNC with the median predictor, a configuration of 10 neurons consistently yielded optimal performance across all scenarios. This architecture provided stable, low-error estimates regardless of the distortion type or intensity, consistently outperforming the direct

NNCC model. The resulting mean error remained around 2 mm and did not exceed 7 mm even in the most challenging case of strong barrel distortion.

It is important to note that RBNC and NNCC share the same input structure—that is, the spatial coordinates in the source system—so both methods may be affected when control points become spatially clustered or overlapping. However, RBNC is not trained to reproduce the full coordinate transformation, but rather to model the residuals left by an initial geometric transformation. This distinction makes its learning task more localized and less complex. Even under these spatial constraints, the residuals often preserve structured, coherent patterns—especially in strong distortion scenarios (Fig. 7)—which serve as informative learning targets. As a result, the network remains capable of extracting meaningful correction functions even in regions where input variability is reduced. This is particularly relevant because the residuals exhibit spatial correlation: in distorted scenarios, especially near the grid boundaries, residual vectors tend to follow similar directions and magnitudes. This coherent structure compensates for the lack of diversity in the input space and allows the RBNC to produce robust correction models even in overlapping regions.

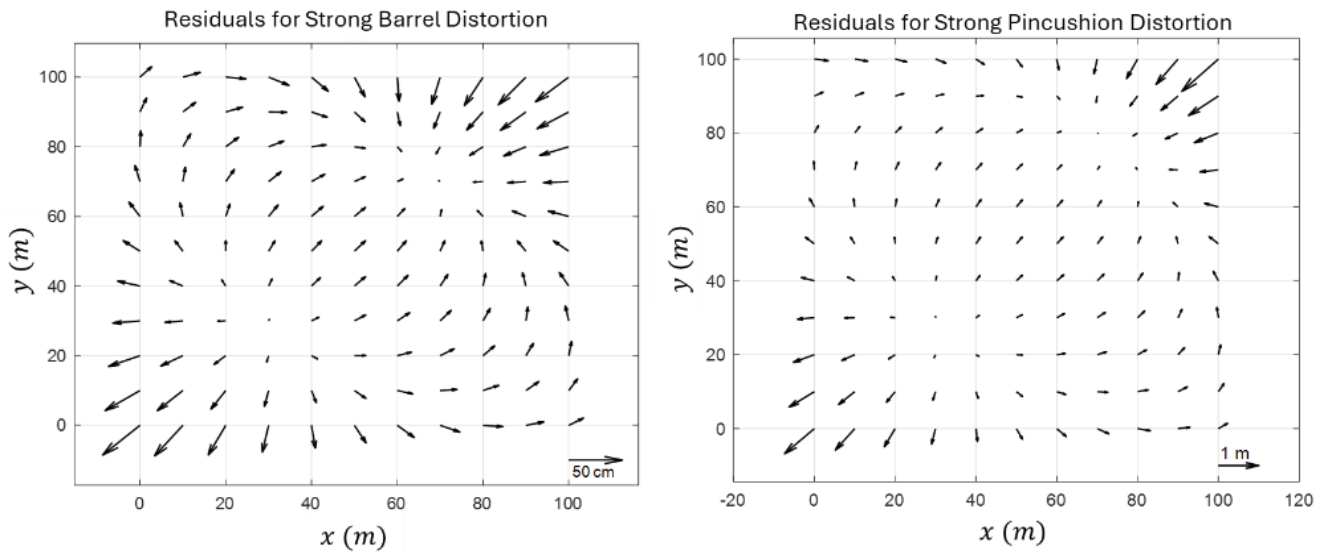

**Figure 7:** Residual vector patterns for barrel (left) and pincushion (right) distortions under strong distortion level.

To complement the quantitative analysis, Figures 8 and 9 show the spatial distribution of the resulting errors under barrel and pincushion distortions, respectively, across three levels of intensity (strong, moderate, and weak, from top to bottom). Each column corresponds to one of the methods—RBNC-median, NNCC-median, and GeoTransf—using the optimal neuron configuration previously identified for each case.

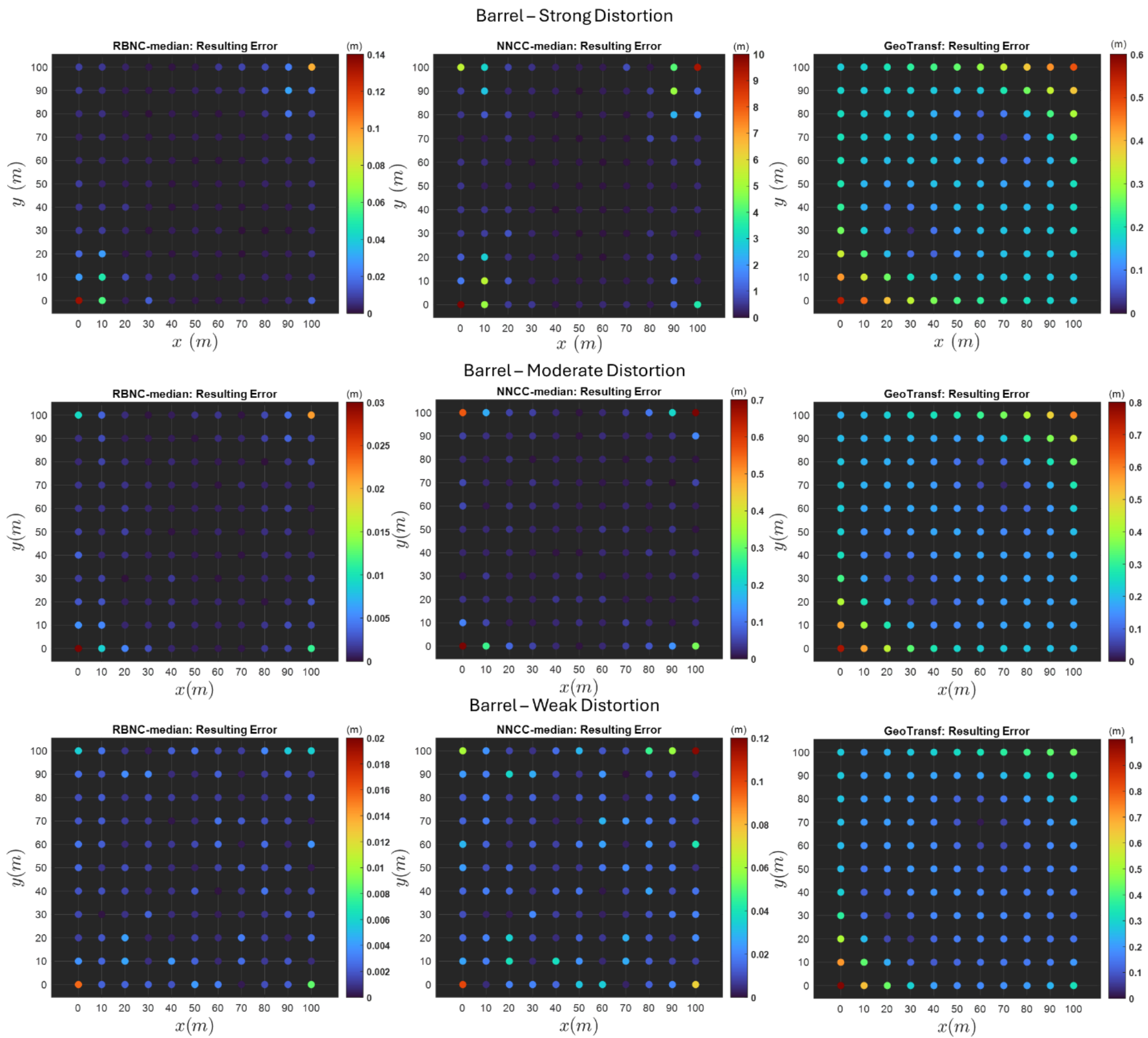

**Fig. 8.** Spatial distribution of resulting errors for the RBNC-median, NNCC-median, and GeoTransf methods under barrel distortion: strong (top), moderate (middle), and weak (bottom).

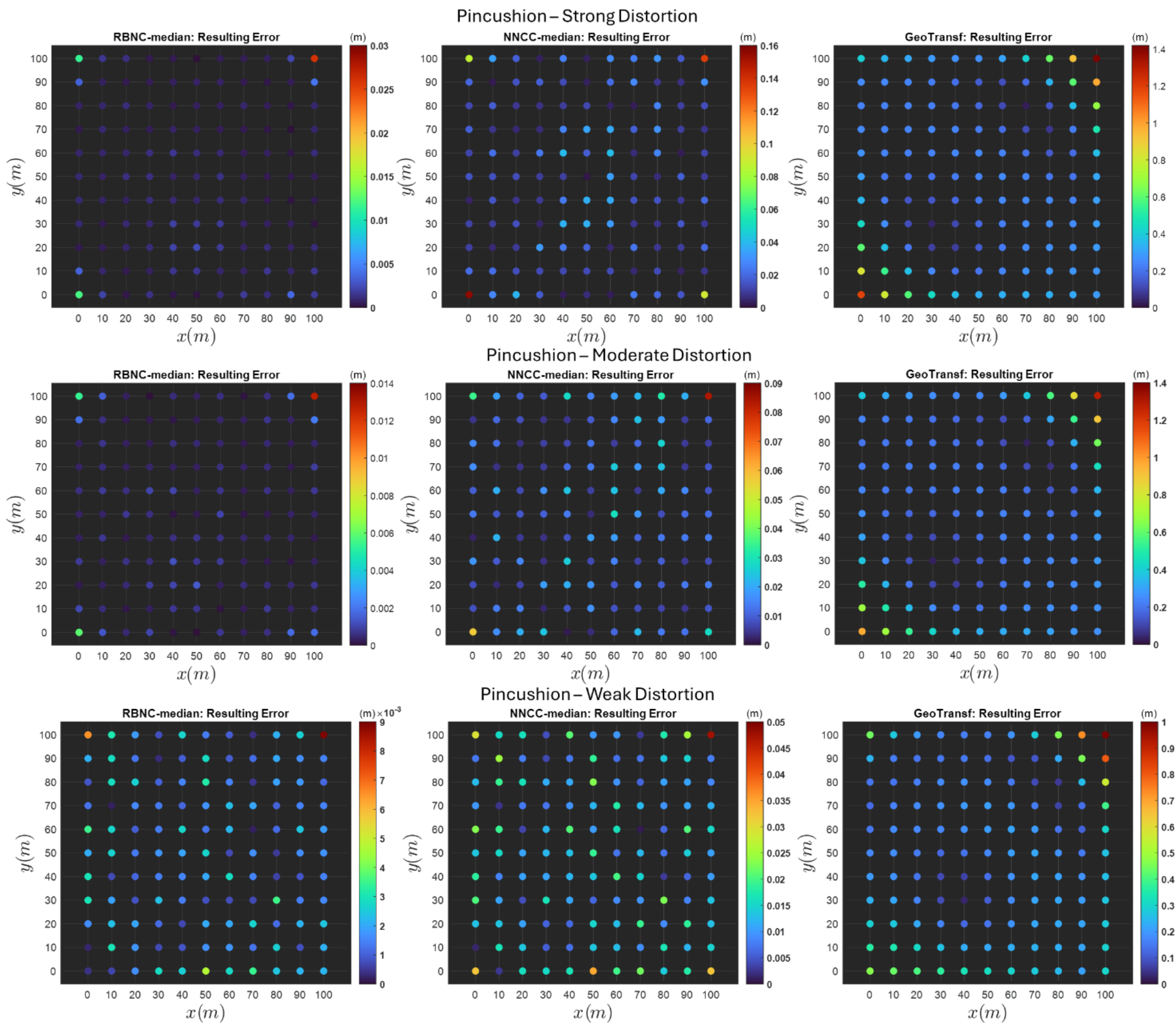

Fig. 9. Spatial distribution of resulting errors for the RBNC-median, NNCC-median, and GeoTransf methods under pincushion distortion: strong (top), moderate (middle), and weak (bottom).

A consistent pattern emerges across all evaluated scenarios: the RBNC-median method yields the lowest and most spatially uniform errors. Even under severe distortions, deviations remain confined to a few centimetres, predominantly near the grid boundaries, while the central region exhibits minimal residuals. As the distortion intensity decreases, the error magnitudes drop further—reaching sub-centimetre levels—and the solution becomes essentially uniform across the entire grid.

In contrast, the NNCC-median method shows significant sensitivity to distortion severity. Under strong barrel distortion, large errors—exceeding several tens of meters—accumulate at the grid edges, primarily due to overlapping input coordinates (see Fig. 6). Although the performance improves with weaker distortions, NNCC-median consistently performs worse than RBNC-median across all tested conditions.

The GeoTransf model, in turn, results in scattered and inconsistent error distributions regardless of distortion intensity. Even in mild scenarios, the method fails to produce notable improvements, which highlights its inherent limitation in capturing complex non-linear effects due to insufficient parameterization.

The effect of edge-induced degradation is particularly visible in the strong barrel distortion case, where all methods exhibit elevated errors at the periphery of the grid. This behaviour aligns with the radial nature of the applied distortion, which intensifies toward the borders and increases the discrepancy between observed and modelled positions. While all methods are affected, RBNC-median displays superior robustness. Its errors remain substantially lower and more spatially stable than those observed in the NNCC-median and GeoTransf solutions, even in regions with clustered or poorly distributed control points. This suggests that RBNC is more effective in modelling steep non-linear distortion gradients, particularly in geometrically unfavourable zones.

These findings confirm that RBNC not only offers good correction for unmodelled distortions but also maintains resilience in degenerate geometric configurations, such as those involving overlapping or edge-clustered control points. This robustness arises from the use of residuals as the learning target, which retain essential information about the distortion field independently of the spatial arrangement of the inputs.

## 3.3 Field Experiment with Real Data

This experiment aimed to evaluate the proposed method under real-world conditions, with the goal of transforming image-based coordinates, denoted by $(C, L)$, into the corresponding object-space coordinates, denoted by $(u, v)$. In this context, the image coordinates $(C, L)$ represent the source system, while the object-space coordinates $(u, v)$, measured with a total station, were referenced to the target system (also referred to as *ground truth*). The measurements were taken from signalized points fixed to a wall of the power generator house located on the Monte Carmelo campus of the Federal University of Uberlândia (UFU), in Minas Gerais, Brazil (Figure 10).

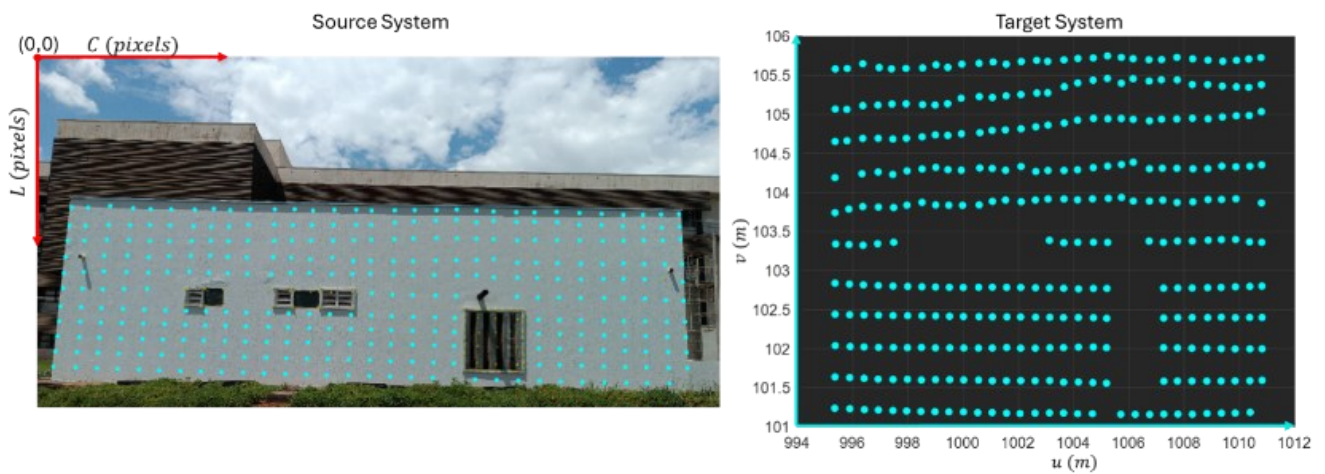

**Fig. 10.** Coordinates registered in the image system (source system) $(C, L)$ and corresponding coordinates obtained in the object-space system (target system) $(u, v)$.

The dataset consisted of 313 homologous points. The image was acquired using a low-cost smartphone nonmetric camera, with dimensions of 12,000 × 9,000 pixels. As highlighted by Karara and Abdel Aziz (1974) [18], non-metric cameras, although not originally designed for photogrammetric purposes, can still be used effectively for spatial measurements provided that appropriate calibration procedures are applied. The points in image coordinate system were manually read and extracted using the *cpselect tool* in MATLAB. The corresponding object coordinates were obtained in two independent field campaigns using a total station, which provided high-precision measurements with a standard-deviation of $\sigma = 2mm$, which guarantees a reliable *ground truth*.

To assess the robustness and generalization ability of the proposed RBNC, NNCC, and GeoTransf, a series of experiments was conducted using this real dataset. The analysis considered not only the number of neurons in the neural networks but also the number and sampling strategy of control points. For this purpose, the original dataset, composed of 313 control points, was deliberately reduced to simulate real scenarios with limited control point availability. The following subsampling strategies were adopted:

A) 50% of the original points selected randomly (Figure 11a);
B) 36 points selected systematically (Figure 11b);
C) 31 points selected randomly (Figure 11c); and
D) 21 points selected systematically (Figure 11d).

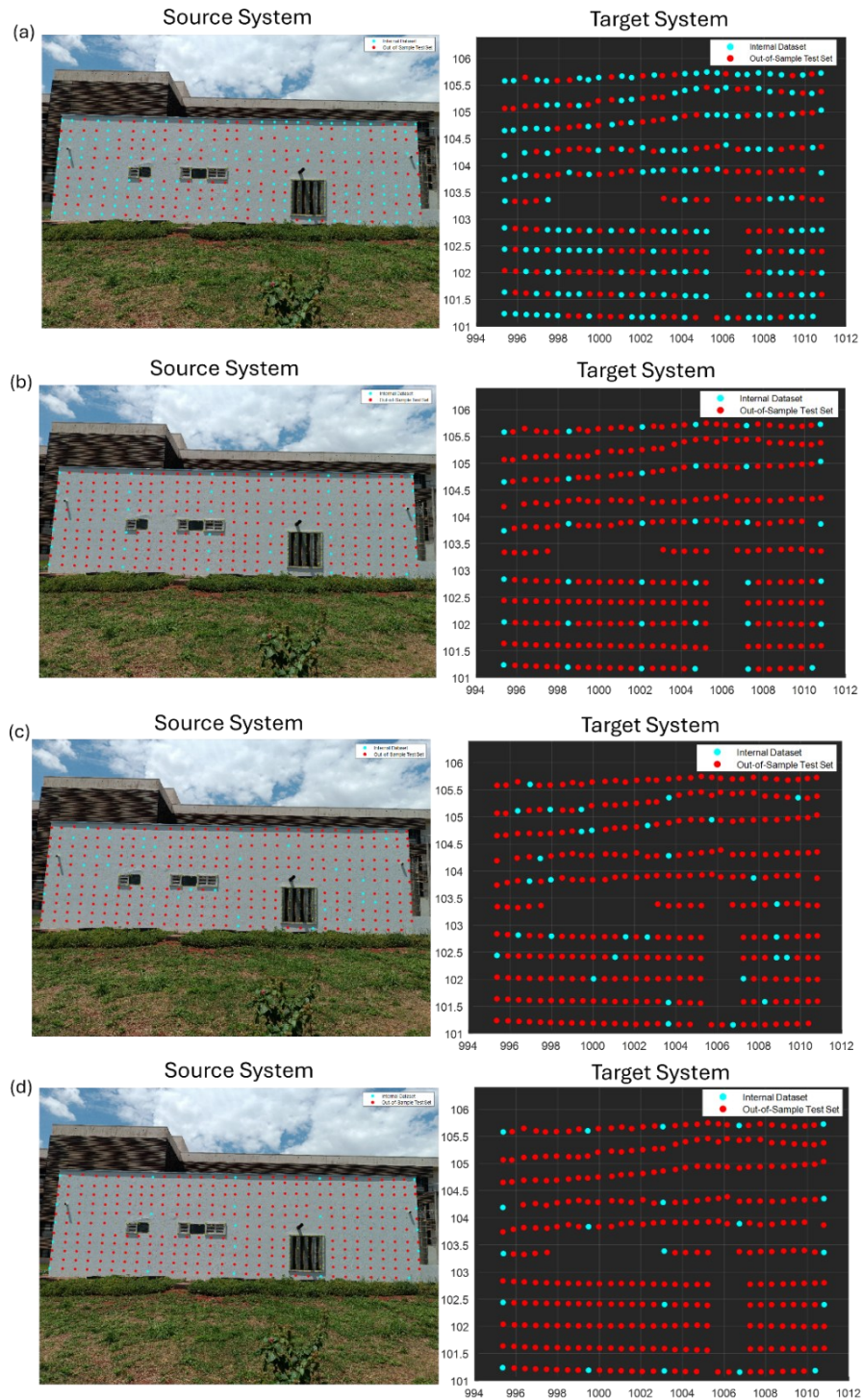

**Fig. 11**. Visualization of the point subsampling strategies: (a) 50% random, (b) 36 systematic, (c) 31 random, (d) 21 systematic. Cyan = Internal Validation, Red = *External Validation*.

Importantly, the remaining points—those not included in the subsample—were entirely excluded from training and internal validation. These held-out samples, referred to as the *out-of-sample test set*, were reserved for external validation to guarantee an unbiased assessment of model performance. For example, in case (A), 157 points were used for RLOOCV (cyan circle in Figure 11a), while the other 156 served as the out-of-sample test set (red circle in Figure 11a); in case (B), 277 were used for testing (red circle in Figure 11b); in case (C), 282 (red circle in Figure 11c); and in case (D), 292 (red circle in Figure 11d). These points were treated as entirely new observations and were not used at any stage for model selection, such as choosing the number of neurons or adjusting other neural network hyperparameters. Likewise, they did not participate in the estimation of transformation parameters for the GeoTransf method.

The same geometric transformation model described in Equations (15) and (16) was used to estimate the transformation parameters. For both RBNC and NNCC, the evaluation procedure followed the same protocol used in the simulated experiment described in Section 3.1.

The RLOOCV and LOOCV procedures were applied to the internal dataset, shown in cyan in Fig. 11. The external dataset (*out-of-sample test set*), shown in red in Fig. 11, was used to evaluate the performance of the methods under real operational conditions by computing the prediction errors using Equation (25) for GeoTransf and Equation (26) for RBNC and NNCC.

## 3.4 Results for the Field Experiment

The analysis of the real field experiment is organized into two parts: internal validation and external validation. First, the internal validation results—derived from the internal dataset using the RLOOCV strategy—are summarized in Table 1.

As expected, the classical GeoTransf method exhibited a structurally limited capacity to handle complex distortions, with mean errors consistently ranging between approximately 9 to 13 mm across all sample sizes. Notably, the performance of GeoTransf remained unaffected by the size data, as the method lacks mechanisms to model unaccounted distortions or adapt to varying sample densities. While the error dispersion was relatively stable, with standard deviations between 3.8 and 6.8 mm, the method systematically underperformed when compared to the neural network-based approaches. This highlights its inability to capture the nonlinearities inherent in the transformation between the image-based and object-space coordinates, which becomes particularly critical in scenarios involving geometric distortions.

**Table 1.** Summary of methods performance for different sample sizes for internal validation.

| Sample Size | Method | Optimal Neurons | Best Estimator | Mean Error (mm) | Std. Dev. (mm) | Notes |
|---|---|---|---|---|---|---|
| 21 Systematic | GeoTransf | - | Median | 11.7 | 6.0 | High error, no distortion correction |
| 21 Systematic | NNCC | 4 | Median | 9.2 | 6.3 | Improvement over GeoTransf, sensitive to sparsity |
| 21 Systematic | RBNC | 10 | Median | 7.6 | 4.7 | Best performance, robust under low samples |
| 31 Random | GeoTransf | - | Median | 9.1 | 6.8 | Same limitation, random sparse points |
| 31 Random | NNCC | 8 | Median | 4.0 | 5.0 | Significant improvement, sensitive to neuron count |
| 31 Random | RBNC | 10 | Median | 3.6 | 4.6 | Best performance, stable under ~4 mm |
| 36 Systematic | GeoTransf | - | Median | 12.9 | 3.8 | Poor, no adaptation to distortion |
| 36 Systematic | NNCC | 8 | Median | 4.2 | 2.2 | Good with tuned neurons |
| 36 Systematic | RBNC | 10 | Median | 2.3 | 1.8 | Best, consistent performance under ~2 mm |
| 157 (50%) Random | GeoTransf | - | Median | 9.9 | 4.1 | No improvement with larger sample |
| 157 (50%) Random | NNCC | 16 | Median | 1.4 | 0.8 | Competitive with RBNC at this density |
| 157 (50%) Random | RBNC | 10 | Median | 1.4 | 0.8 | Maintains robustness, within ~2mm |
| 313 (100%) Full dataset | GeoTransf | - | Median | 9.7 | 3.9 | Structural limitation |
| 313 (100%) Full dataset | NNCC | 16 | Median | 1.2 | 0.8 | Best NNCC performance, close to RBNC |
| 313 (100%) Full dataset | RBNC | 10 | Median | 1.2 | 0.7 | Benchmark performance |

The NNCC method demonstrated a clear advantage over GeoTransf, particularly when the sample size allowed for a sufficient diversity of input patterns. With the complete dataset (313 points), NNCC achieved its best performance with 16 neurons, resulting in a mean error of 1.2 mm and a standard deviation of 0.8 mm—this outcome matched the RBNC method and remained well within the reference uncertainty of 2 mm. Similar results appeared in the 157-point subsample, where NNCC maintained high accuracy.

However, NNCC's performance proved to be highly sensitive to sample size. For smaller datasets, especially with the systematic selection of 21 points, performance deteriorated noticeably. Despite optimization to 4 neurons and the adoption of the median estimator, the mean error increased to 9.2 mm, and the dispersion widened to 6.3 mm. The method's vulnerability to sample sparsity and its dependence on careful tuning of the network complexity became evident. Nonetheless, it is important to note that even under these challenging conditions, NNCC still managed to outperform GeoTransf, underlining its potential when the network architecture is properly adjusted. Across all cases, the median estimator consistently delivered better performance than the mean, confirming its superior robustness against outliers and its suitability for scenarios with higher error variance (see the Supplementary Material).

The RBNC method emerged as the most reliable and robust approach across all tested scenarios. Crucially, the configuration of 10 neurons—previously identified in simulated experiments—proved optimal in every real-world case, regardless of the sample size or selection strategy. Remarkably, RBNC consistently maintained mean errors within or very close to the expected reference uncertainty of 2 mm.

For the most reduced sample of 21 systematically selected points, RBNC achieved a mean error of 7.6 mm, substantially lower than both GeoTransf (11.7 mm) and NNCC (9.2 mm). With intermediate samples, such as the 36-point systematic set and the 31-point random selection, RBNC sustained mean errors of 2.3 mm and 3.6 mm, respectively, demonstrating strong resilience to variations in data sparsity and distribution. Notably, in the denser datasets (157 and 313 points), RBNC performed on par with NNCC, both reaching errors around 1.2–1.4 mm, comfortably within the target accuracy range.

Following the internal validation and optimal model selection for each subsampling strategy, the models generated from 157, 36, 31, and 21 points were applied to their corresponding out-of-sample test sets to evaluate generalization under entirely unseen conditions. Specifically, these configurations were evaluated on 156, 277, 282, and 292 external points, respectively. These test sets were strictly excluded from all stages of model development, including RLOOCV and LOOCV loops, neuron selection, and parameter estimation, ensuring an unbiased evaluation.

The results for external dataset, visually summarized in Fig. 12, reveal a clear trend across all methods for varying sample sizes and selection strategies. The marker colors indicate the magnitude of the resulting errors: **Blue** for errors ≤ 3 mm, **Green** for errors > 3 mm and ≤ 4 mm, **Cyan** for errors > 4 mm and ≤ 5 mm, **Yellow** for errors > 5 mm and ≤ 6 mm, **Magenta** for errors > 6 mm and ≤ 8 mm, **Red** for errors > 8 mm and ≤ 10 mm, and **Black** for errors > 10 mm (1 cm).

As expected, GeoTransf consistently exhibited the highest error rates, with widespread deviations exceeding 1 cm across the wall surface, which reflects its inability to capture the nonlinear distortion effects inherent in the transformation task. NNCC, in turn, delivered marked improvements over GeoTransf, particularly in the models generated from larger sample sizes. For example, in the 157-point random and 36-point systematic configurations, NNCC produced competitive results, with most points concentrated below the 6 mm threshold. However, performance declined significantly in the most reduced dataset of 21 systematically selected points, where larger residual errors appeared more frequently, highlighting NNCC's sensitivity to both sample size and distribution.

RBNC consistently demonstrated superior performance across all external validations. For the generated models with 157-point random and 36-point systematic datasets, RBNC effectively controlled the distortions, maintaining most predictions below 5 mm and preserving overall spatial coherence. Notably, for the RBNC method, the use of systematic sampling consistently resulted in superior performance compared to random sampling strategies. In the case of 36 systematic points, RBNC sustained its robustness, delivering predominantly 3 mm errors, reinforcing its adaptability to more structured control point layouts. Nonetheless, in the most challenging scenario of 21 systematically selected points, RBNC, despite remaining the best-performing method, exhibited noticeable error increase. The mean error reached approximately 7.6 mm, with clusters of points exceeding 1 cm, indicating that even the residual-based approach faces limitations when constrained by extreme data sparsity.

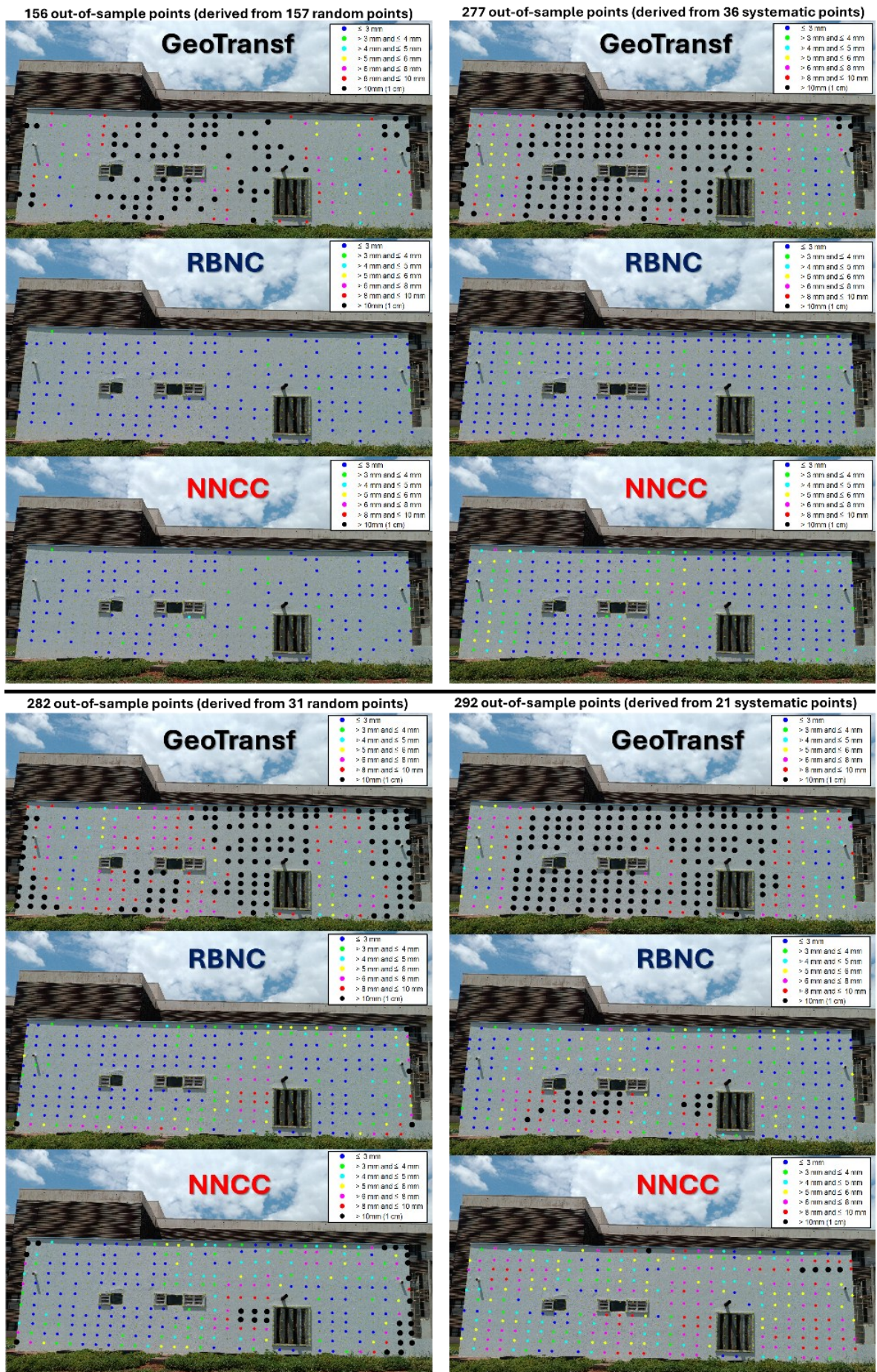

**Fig. 12.** Spatial distribution of resulting errors for the out-of-sample test sets under different control point configurations.

These findings emphasize the importance of sufficient and well-distributed control points to fully leverage the capabilities of neural network-based correction models in practical applications. These findings for RBNC are further corroborated by the residual vector field analysis (Fig. 13), which reveals the systematic distortion patterns that persist after the application of the geometric transformation.

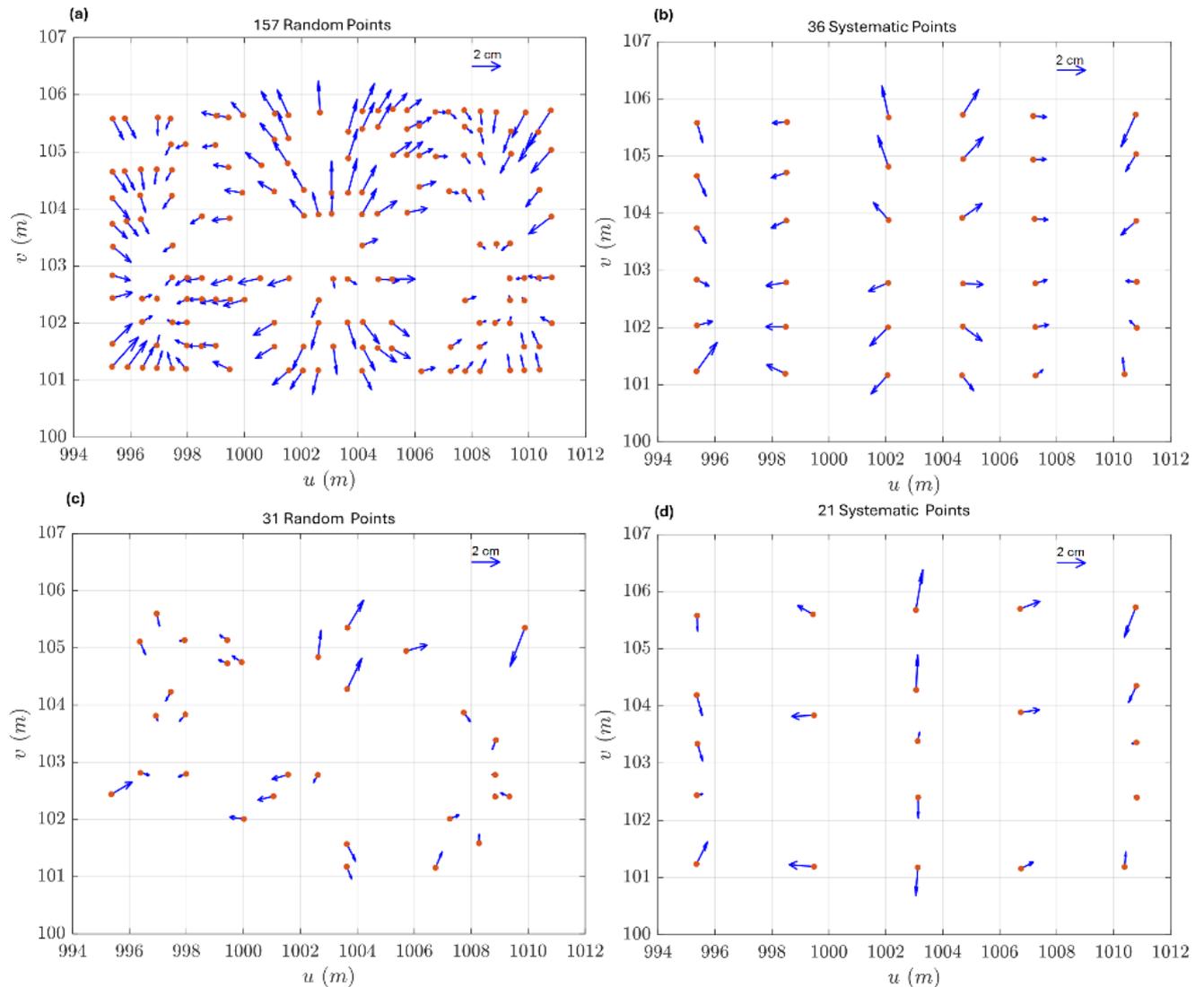

**Fig. 13**. Residual vector fields after geometric transformation, by sample size and selection strategy: (a) 157 random points; (b) 36 systematic points; (c) 31 random points; (d) 21 systematic points.

As previously observed in the simulation study, the residuals preserve a structured and meaningful pattern even in real-world conditions, enabling the neural network in the RBNC approach to effectively model the uncorrected distortions. This stepwise process — first applying the analytical transformation, followed by residual modelling — allowed the RBNC to capture the complex geometric distortions present in the field data, substantially improving predictive accuracy.

Overall, the external validation confirms the robustness of the RBNC approach, even under challenging constraints, while also illustrating the inherent trade-offs associated with neural networks when control point availability becomes critically sparse. These results, aligned with the trends observed in the internal validation, further consolidate RBNC as a reliable solution for coordinate transformation tasks under variable field conditions.

## 4. CONCLUSION

This study confirms that neural network-based correction models offer clear advantages over classical geometric transformations in handling unmodeled distortions in coordinate transformation tasks. The proposed RBNC approach consistently outperforms both the direct neural converter (NNCC) and the analytical model (GeoTransf) across all tested scenarios. Rather than depending on predefined analytical distortion models, RBNC takes advantage of the structured residual patterns left by the geometric transformation to autonomously model and correct complex, spatially dependent distortions. This approach emerges as an effective alternative to conventional modelling techniques, particularly when explicit parametrization of distortions becomes impractical.

The analysis of internal and external validations confirms that RBNC consistently maintains accuracy within the reference uncertainty of the target system, even under challenging scenarios with severely reduced and systematically sampled control points. This robustness highlights the method's capacity to reliably model and correct distortions, regardless of sample constraints. By contrast, although NNCC performs competitively in denser datasets, it becomes unsuitable in sparse configurations, especially when overlapping input points occur — a condition that critically undermines its generalization capacity. The accumulation of nearly identical input attributes compromises the neural network's ability to model distinctive patterns, rendering NNCC impractical under these circumstances. Meanwhile, the GeoTransf method predictably exhibits the highest error levels, reflecting its inherent structural limitations and inability to accommodate complex distortion effects.

Crucially, the residual vector field analysis reveals that RBNC benefits from the structured patterns inherent in the residuals, which enables effective correction even when control points are sparse and unevenly distributed. Systematic sampling strategies further enhance the network's ability to generalize, reinforcing the importance of thoughtful control point distribution in practical applications.

Collectively, these findings establish RBNC as a robust and reliable solution for coordinate transformations in scenarios where traditional models fall short, particularly in the presence of complex distortions and constrained control point availability. The method's simplicity—requiring only a modest neural architecture of 10 neurons—underscores its practicality for real-world deployment, offering a balance between computational efficiency and predictive accuracy.

In future work, the proposed method will be compared with other transformation approaches that include distortion modelling between reference frames, such as the *Thin Plate Splines* technique. This method has shown promising results in geodetic coordinate transformations involving systematic distortions, as demonstrated by Magna Júnior, Camargo, and Galo (2014) in the transformation between SAD69 and SIRGAS2000 [3].

## CRediT authorship contribution statement

**Vinicius Francisco Rofatto:** Conceptualization, Data curation, Formal analysis, Methodology, Validation, Visualization, Writing – original draft, Writing – review & editing, Supervision, Funding acquisition. **Luiz Felipe Rodrigues de Almeida:** Data Acquisition, Formal analysis, Methodology, Writing – review & editing. **Ivandro Klein:** Methodology, Formal analysis, Writing – review & editing. **Marcelo Tomio Matsuoka:** Methodology, Formal analysis, Writing – review & editing. **Mauricio Roberto Veronez:** Writing – review & editing, Supervision, Funding acquisition. **Luiz Gonzaga Da Silveira Junior:** Writing – review & editing, Supervision, Funding acquisition.

## Declaration of competing interest

The authors declare that there are no conflicts of interest.

## Acknowledgments

The authors would like to thank the Laboratory of Topography and Geodesy at the Monte Carmelo Campus of the Federal University of Uberlândia (UFU). This project was funded by the National Council for Scientific and Technological Development – CNPq – Grant No. 421278/2023-4.

## Appendix A. Supplementary data
Rofatto, V.; Almeida, L. F. R.; Klein, I.; Matsuoka, M. T. (2025). Residual-based Neural Network for Unmodeled Distortions in Coordinate Transformation [Data set]. Zenodo. https://doi.org/10.5281/zenodo.15226585